\documentclass[12pt]{iopart}

\usepackage{color}

\usepackage{graphicx}
\begin{document}

\title[]{Efficient Green Emission from Edge States in  Graphene Perforated by Nitrogen Plasma Treatment}

\author{N N Kovaleva$^{1*}$, D Chvostova$^{2}$, Z Pot\r u\v cek$^{2,3}$, H D Cho$^4$, \mbox{Xiao Fu$^5$}, L Fekete$^2$, J~Pokorny$^2$, Z~Bryknar$^3$, K I Kugel$^{6,7}$, \mbox{A Dejneka$^2$}, \mbox{T W Kang$^5$}, Gennady~ N~ Panin$^{5,8}$ and \mbox{F V Kusmartsev$^{9,10,11*}$}}

\address{$^1$ Lebedev Physical Institute, Russian Academy of Sciences, Moscow 119991, Russia}
\address{$^2$ Institute of Physics, Academy of Sciences of the Czech Republic, Prague, 18221 Czech Republic}
\address{$^3$ Czech Technical University, Prague, 12000 Czech Republic}
\address{$^4$ Quantum-Functional Semiconductor Research Center, Dongguk University-Seoul, Seoul 04623, Korea}
\address{$^5$ Nano Information Technology Academy, Dongguk University-Seoul, Seoul 04623, Korea}

\address{$^6$ Institute for Theoretical and Applied Electrodynamics, Russian Academy of Sciences, Moscow  125412, Russia}

\address{$^7$ National Research University Higher School of Economics, Moscow  101000, Russia}

\address{$^8$ Institute of Microelectronics Technology and High-Purity Materials, Russian Academy of Sciences, Chernogolovka 142432, Russia}

\address{$^9$ Department of Physics, Loughborough University, Loughborough LE11 3TU, UK}
\address{$^{10}$Microsystal $\&$ Terahertz Research Center, Chengdu 610200, Sichuan, China}
\address{$^{11}$ITMO University, St. Petersburg 197101, Russia}
\eads{\mailto {corresponding authors: n.kovaleva@lboro.ac.uk} and\mailto{ f.kusmartsev@lboro.ac.uk}}



\begin{abstract}
Plasma functionalization of graphene is one of the facile ways to tune its doping level without the need for wet chemicals making graphene photoluminescent. Microscopic corrugations in the two-dimensional structure of bilayer CVD graphene having a quasi-free-suspended top layer, such as graphene ripples, nanodomes, and bubbles, may significantly enhance local reactivity leading to etching effects on exposure to plasma. Here, we discovered that bilayer CVD graphene treated with nitrogen plasma exhibits efficient UV-green-red emission, where the excitation at 250 nm leads to photoluminescence with the peaks at 390, 470, and 620 nm, respectively.  By using Raman scattering and spectroscopic ellipsometry, we investigated doping effects induced by oxygen or nitrogen plasma on the optical properties of single- and bilayer CVD graphene. The surface morphology of the samples was studied by atomic force microscopy. It is revealed that the top sheet of bilayer graphene becomes perforated after the treatment by nitrogen plasma. Our comprehensive study indicates that the dominant green emission is associated with the edge defect structure of perforated graphene filled with nitrogen. The discovered efficient emission appearing in nitrogen plasma treated perforated graphene may have a significant potential for the development of advanced optoelectronic materials.
\end{abstract}


%
%
%
%

\section{Introduction}
Graphene composed of sp$^2$ bonded carbon atoms arranged in a two-dimensional honeycomb lattice structure exhibits fascinating and exotic properties, being the thinnest and strongest ever known material, impermeable to gases, showing record stiffness, etc. Graphene is a zero-bandgap semiconductor, where Dirac electrons propagating in the lattice effectively lose their mass, which also leads to outstanding electrical characteristics such as high electron mobility and conductivity. These outstanding transport properties represent the most explored aspect of graphene physics attractive for its application in various electronic devices [1-3]. However, due to the lack of a bandgap in the electronic spectrum of graphene, the possibility of observing light emission or photoluminescence (PL) from the high-quality graphene is highly unlikely. The PL property of graphene can be derived by tuning the bandgap by doping with various reactive functional groups. If subjected to functionalization by fluorine [4],  oxygen [5,6],  nitrogen [7], or hydrogen [8] graphene is highly promising for high-performance electronics and optoelectronics [5,9,10]. In addition, reducing the size of a graphene sheet down to a scale of 1 nm, one can obtain isolated graphene clusters, or graphene quantum dots (GQDs), which may have large bandgaps due to the finite number of atoms in the clusters tuned by the GQD size [11-15]. GQDs are considered to be the next generation carbon-based nanomaterials, which have attracted much attention in bio-imaging, light-emitting, and photovoltaic applications [16-20]. 

Functionalization of graphene and GQDs can be done by chemical and physical methods. In 1958 Hummers and Offeman developed a basic chemical oxidation method by reacting graphite with a mixture of potassium permanganate (KMnO$_4$) and sulfuric acid (H$_2$SO$_4$) [21]. Most commonly, graphene oxide (GO) sheets are exfoliated from graphite flakes, and as-synthesized GO is insulating [19,20]. GO prepared by the chemical functionalization contains various reactive oxygen functional groups, primarily epoxides (C$_2$O) and hydroxyls (C-OH) on the basal plane, as schematically shown in figure 1(a), with a very low amount of carboxylic acids (HCO$_2$) and carbonyl (in the form of water molecule H$_2$O) at the sheet edge (not shown). In contrast to as-synthesized GO, partially oxidized graphene, where the main oxygen functional groups are in the form of epoxides, is conductive [22,23]. In turn, graphene functionalization with nitrogen may lead to the formation of N-graphene (NG), where three bonding configurations with carbon atoms, including pyridinic N, pyrrolic N, and quaternary N (or graphitic N) dominate inside the pristine graphene lattice [24], as schematically shown in figure 1(b). Specifically, pyridinic N bonds with two carbon atoms at the edges or defects of graphene and in quaternary N nitrogen atoms substitute three carbon atoms in the hexagonal ring, where they are sp$^2$ hybridized, whereas pyrrolic N is bonded into a five-membered ring and is sp$^3$ hybridized, which induces a local buckling. Nitrogen doped graphene can be prepared by the modified Hummers method [25] and the following ammonia (NH$_3$) heat treatment process. The N content (of about 2.8 at.\%) in the chemically synthesized NG is relatively low [26]. The higher N content up to 8.9 at.\% is reported for NG obtained in direct CVD synthesis [27]. However, the methods for the production of large-area N-graphene are still lacking. Physical methods of functionalization of graphene include plasma treatment (as post treatment), which is one of the facile ways to tune the intrinsic properties of graphene without using wet chemicals to make graphene photoluminescent. However, most plasma treatments lead to surface doping only. It was shown that PL emission can be induced in a mechanically exfoliated single-layer graphene flake exposed to oxygen/argon (1:2) RF plasma treatment (with the plasma strength: 0.04 mbar and 10 W) during an exposure time of one hour [4]. The plasma induced PL emission was tempted to be interpreted as coming from the small formed sp$^2$ domains (or GQDs) with an average size of $\sim$1 nm. However, finally, its origin was referred to CO-related defects created by the plasma treatment. The multilayer graphene remained nonluminescent following the plasma treatment [4]. When graphene is placed into nitrogen plasma, carbon atoms can partly be replaced by nitrogen atoms, and this occurs more probably at the defects and edges of graphene grains or flakes. Then, the low doping level in processing with nitrogen plasma may be attributed to the low defect concentration in the high quality graphene. According to the earlier studies, the reached N content, which can be controlled by the plasma strength and exposure time, varies from 3 to 8.5 at.\% [28,29]. Other nitrogen doping methods include N-doping of graphene, for example, through electrothermal reactions with NH$_3$ [30], by performing the arc discharge of graphite electrodes in the presence of H$_2$, He, and NH$_3$ [31], by controlled electrochemical deposition [32], etc.
\begin{figure}
\hspace{2.50cm}
\includegraphics*[width=130mm]{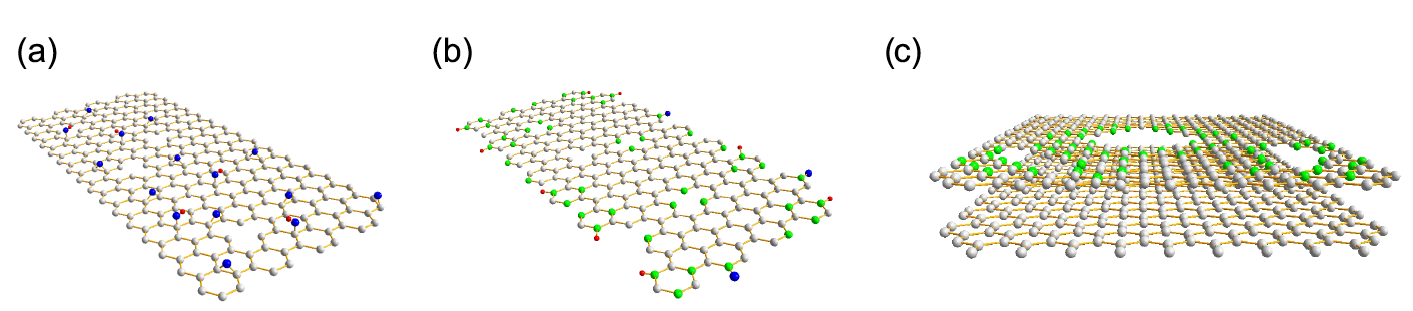}
\caption{Doping of single-layer graphene with (a) oxygen- and (b) nitrogen-related functional groups. (c) Scheme of the perforated graphene layer produced by nitrogen plasma treatment of bilayer CVD graphene.}
\label{Fig1}
\end{figure}

Here, we concentrate our attention on the morphology aspects in plasma treatment of graphene, which have not yet been fully explored. It is well known that freely suspended graphene has ``intrinsic'' ripples, nanodomes, and bubbles [33-39]. The existing microscopic corrugations have nanosized lateral dimensions and a height displacement of about 0.7 to 1 nm. The bigger ripples may attain even 2-3 nm in height resulting, in particular, in strain-induced local conductance modulations [40,41]. The curved regions leading to local bonding distortions could result in some sp$^3$ orbital character, as well as in the $\pi$--orbital misalignment, which is expected to significantly increase the local reactivity in the process of graphene functionalization [42--44]. Then, microscopic corrugations in the two-dimensional structure of freely suspended graphene, as well as those appearing in the process of transfer of CVD-grown graphene sheets to a substrate [45], can give rise to enhanced etching effects occurring on exposure to plasma. Since the graphene rippling can be strongly suppressed by the interfacial van der Waals interactions in single-layer CVD graphene supported by substrate, in bi- or few-layer CVD graphene, where the adjacent graphene sheets are only weakly interacting due to the enlarged interlayer distance, the morphology aspects [34-37] may become important. This may also lead to a strong dependence of the plasma etching kinetics on the number of layers [46].

Herein, by using several optical experimental techniques, namely, Raman scattering (RS), spectroscopic ellipsometry (SE), and photoluminescence PL, we studied the optical properties of the single- and bilayer graphene (SLG and BLG) samples exposed to oxygen (with the plasma strength: 0.06 mbar and 20 W) and nitrogen (0.34 mbar and 20 W) plasmas with an exposure time of 90 s at room temperature. In addition, the AFM technique was applied to investigate the effect of post treatment with oxygen or nitrogen plasma on the surface morphology of SLG and BLG grown by the CVD method and transferred to a SiO$_2$/Si substrate. The Raman spectra of the SLG samples exhibit an equivalent plasma treatment effect identifying a significant introduced doping level, as reported for a mechanically exfoliated SLG flake exposed to oxygen/argon (1:2) RF plasma (0.04 mbar, 10 W, 1 s) [4]. The optical conductivity spectra obtained using  the multilayer modeling of the measured SE response in the effective medium approximation (EMA) indicate that the SLG sheets are only partially functionalized. We observed that PL emission induced in the SLG samples (SLG/SiO$_2$/Si)is relatively weak and most probably can be associated with defective structures containing carbon vacancies and epoxides (for O$_2$ plasma treatment) or pyridinic (for N$_2$ plasma treatment) functional groups. The observed downshifts of the main Raman modes indicate that bilayer CVD graphene has a quasi-free-suspended top layer. According to the SE study, the top BLG sheet treated under the applied oxygen plasma conditions is practically unaffected. In contrast, the strongly decreased extinction coefficient and effective optical conductivity indicate a significantly increased porosity of the top BLG sheet treated with nitrogen plasma. Here, we have discovered efficient UV-green-red emission from the BLG samples (BLG/SiO$_2$/Si)exposed to the nitrogen plasma treatment, where excitation at 250 nm leads to the PL emission with the peaks at 390, 475, and 620 nm. It has been confirmed by the present AFM study that an upper sheet of bilayer CVD graphene has  a complex morphology and the applied nitrogen plasma conditions lead to its partial etching. As a result of partial etching, the top sheet of BLG acquires a perforated structure and a network of graphene edge defect states is formed. In this case, carbon atoms can easily be replaced on a network of graphene edges by nitrogen radicals, as schematically shown in figure 1(c). Our comprehensive study suggests that the dominant green emission may be associated with the edge defect structure of a quasi-free-suspended layer of perforated graphene filled with nitrogen, which determines the observed PL properties.

\section{Experimental}
Graphene was synthesized by chemical vapor deposition (CVD) at 1020$^\circ$C on 25 $\mu$m Cu foil sheets (99.999\% Alfa Aesar, 12 $\times$ 30 cm$^2$) using a mixture of methane and hydrogen. The as-synthesized graphene on copper was spin coated with a thin layer of polymethyl methacrylate (PMMA) for 1 min to produce a 2 $\mu$m PMMA layer. The sample was then annealed at 120$^\circ$C for 10 min to cure the polymeric carrier. The graphene layer on the back side of the copper sheet was removed in oxygen plasma (60 W power for 10 min). The sample was then placed on the surface of the copper etchant CE-100 for etching for 40 min, and then for washing in two DI water baths for 20 min each. Finally, the PMMA supported graphene was transferred one time (single-layer graphene (SLG)) or two times (bilayer graphene (BLG)) onto a SiO$_2$(300 nm)/Si(100) substrate. PMMA was removed in acetone in an ultrasonic bath for 20 min. Graphene was washed with 30\% HCl (60$^\circ$C) for 30 min to remove the residual Fe$^{3+}$ ions of the copper etchant. Samples of SLG and BLG were treated in a plasma unit with an exposure time of 90 s at room temperature in oxygen plasma (0.06 mbar and 20 W) or in nitrogen  plasma (0.34 mbar and 20 W).

Surface morphology of SLG and BLG on SiO$_2$(300 nm)/Si(100) substrates was studied at room temperature by atomic force microscopy (AFM) using an ambient AFM (Bruker, Dimension Icon) in the Peak Force Tapping mode with ScanAsyst Air tips (Bruker; k = 0.4N/m; nominal tip radius 2 nm).

Polarized Raman spectra were recorded using a Renishaw Raman RM-1000 Micro-Raman spectrometer with CCD detection. The measurements were performed in a backscattering geometry at room temperature. The 514.5 nm line of an Ar$^+$ ion laser was focused to a spot size of about 2 $\mu$m. Standard notch filters were applied for Raman shifts $>$ 200 cm$^{-1}$. 

Dielectric function response of the samples of SLG and BLG was investigated in a wide photon energy range of 0.02-8.5 eV by variable-angle spectroscopic ellipsometry using an IR-VASE Mark II Ellipsometer and a J.A. Woollam VUV-VASE Gen II spectroscopic ellipsometer. The room-temperature ellipsometry measurements were performed for at least three incident angles, usually at 65$^\circ$, 70$^\circ$, and 75$^\circ$. Additionally, the corresponding ellipsometry measurements were performed on the blank SiO$_2$(300 nm)/Si(100) substrate. The measured ellipsometric angles,  $\Psi(\omega)$ and $\Delta(\omega)$, were simulated via multilayer dielectric modeling using the WVASE32 software package. The applied dielectric model, in addition to SLG and BLG sheets and the SiO$_2$/Si substrate, included Cauchy layers between the substrate and graphene sheets. From the simulation, the complex dielectric function, $\tilde \varepsilon(\omega)=\varepsilon_1(\omega)+\rm{i}\varepsilon_2(\omega)$, of the studied single- and bilayer CVD graphene was extracted. The optical conductivity, $\sigma_1(\omega)$, was calculated by using the equation  $\sigma_1(\omega)$=$\frac{1}{4\pi}\omega\varepsilon_2(\omega)$.

Room-temperature PL and photoluminescence excitation (PLE) spectra were measured on the samples of SLG and BLG in a right-angle scattering geometry using a setup equipped with an SPM 2 grating monochromator (Carl Zeiss). The PL was excited with the light from a high-pressure Xe lamp filtered through a double-grating Jobin-Yvon DH 10UV monochromator. The emission was detected in the spectral range from 350 nm (3.54 eV) to 870 nm (1.46 eV) with a cooled RCA 31034 photomultiplier (with a GaAs photocathode) operating in the photon-counting mode. The PL emission spectra were taken with a spectral resolution of 6 nm and corrected for the apparatus spectral dependence. The PLE spectra were referred to a constant flux of excitation light over the whole studied spectral range.

\section{Results and Discussion}

To investigate the surface morphology of graphene sheets grown in this study by the CVD method and transferred onto a SiO$_2$/Si substrate one time (for a SLG sample) or two times (for a BLG sample), as well as to examine possible plasma etching effects, we used the AFM approach. The AFM images of a bare SiO$_2$(290 nm)/Si(100) substrate and that of a SLG sample are presented in figures S1(a) and (b) (of the supplemenrtary online information (stack.iop.org/TDM/00/000000/mmedia)). From the acquired AFM scans, the estimated root-mean-square (RMS) roughness of the SiO$_2$/Si substrate is about 0.5 nm, whereas the examined SLG sample demonstrated slightly higher RMS roughness of about 0.6 nm. From inspection of the SLG boundary, the average evaluated SLG height with respect to the SiO$_2$/Si substrate level is about 0.7 nm (see Figure S1(c) and inset). Figures 2(a) and (d) show 3 $\times$ 3 $\mu$m$^2$ AFM images of two BLG samples: that of as-transferred original one and that of exposed to the nitrogen plasma treatment (for more details see the 600 x 600 nm$^2$ zoom area in figures 2(b) and (e)). In comparison to the examined SLG sample, the original BLG sample has noticeably larger RMS roughness of about 1.0 nm. In the AFM image (figure 2(b)) one can see that the upper BLG sheet is represented by the topmost level shown in the yellow contrast and the bottom level depicted in the light-brown and brown contrasts. It follows from the height histogram analysis that the topmost level lies above in height by about 1.2 nm on the average (figure 2(c)). This is consistent with the complex morphology of the structure previously reported for CVD-grown graphene sheets wet-transferred onto a SiO$_2$/Si substrate [45] and can be associated with the presence of nanopockets of intercalated water, residual PMMA and other residues from the used solvents (acetone and HCl, see Section 2) trapped between the adjacent CVD graphene sheets. In addition, in the AFM image one can distinguish the dark-brown and black contrasts, which correspond to one-layer-deep holes existing in the upper BLG sheet. Then, from the height histogram, the average estimated distance between the upper and lower graphene sheets in the BLG sample is about 0.7-1.0 nm. Meantime, the estimated RMS roughness of the BLG sample treated with nitrogen plasma is about 0.5 nm and is well consistent with the RMS roughness of the bare SiO$_2$/Si substrate. Its height histogram is better fitted with two Gaussians (see figure 2(f)). In the AFM image, the upper BLG sheet is represented by the light-brown and brown contrasts corresponding to a fractal-like percolating structure, which occupies about 82\% of the surface area. The rest area is held by one-layer-deep cavities with the transverse lateral size of about 10-20 nm, which are represented by the dark-brown and black contrasts. It is important that the topmost level, which is clearly recognized in the original BLG sample, is absent in the BLG sample treated with N$_2$ plasma. Thus, from the present AFM study we can conclude that (i) the presence of nanopockets of intercalated water or other solvents trapped between the adjacent CVD graphene sheets during wet-transfer to a SiO$_2$/Si substrate leads to the pronounced rippling effects on the upper original BLG sheet and (ii) the nitrogen plasma treatment (with the plasma strength 0.34 mbar and 20 W)  with the exposure time of 90 s leads to etching of the ripples, nanodomes, and bubbles [34-37]. At the same time, according to our AFM study, the applied oxygen plasma conditions (with the plasma strength 0.06 mbar and 20 W) with the exposure time of 90 s lead to weaker etching effect   of the upper BLG sheet at room temperature (see figure S2(a-c)). 
\begin{figure}
\begin{center}
\hspace{2.50cm}
\includegraphics*[width=130mm]{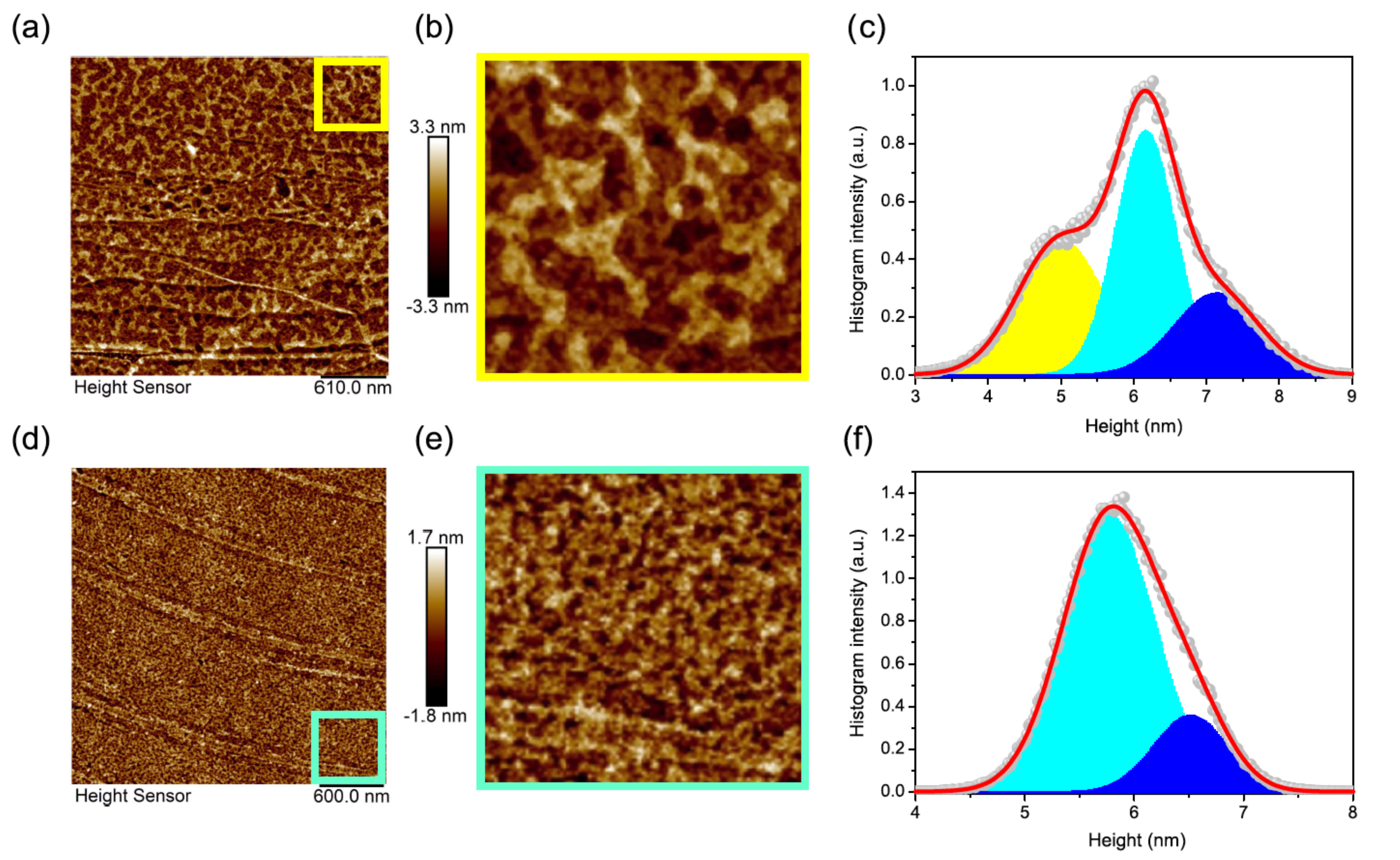}
\end{center}
\caption{AFM images (a,b) of as-transferred BLG/SiO$_2$(300 nm)/Si(100)  and (d,e) of N$_2$- plasma treated BLG/SiO$_2$(300 nm)/Si(100). (a,d) The scan size is 3 $\times$ 3 $\mu$m$^2$. (b,e) Zoom into the marked area of ~600 $\times$ 600 nm$^2$. (c) Fit image with three Gaussian functions (xc$_1$=5.00 nm, w$_1$=1.2 nm; xc$_2$=6.16 nm, w$_2$=0.8 nm; xc$_3$=7.14 nm, w$_3$=1.17 nm). (f) Fit image with two Gaussian functions (xc$_1$=5.77 nm, w$_1$=0.85 nm; xc$_2$=6.52 nm, w$_2$=0.68 nm).}
\label{Fig2}
\end{figure}

\begin{figure}
\begin{center}
\hspace{2.50cm}
\includegraphics*[width=80mm]{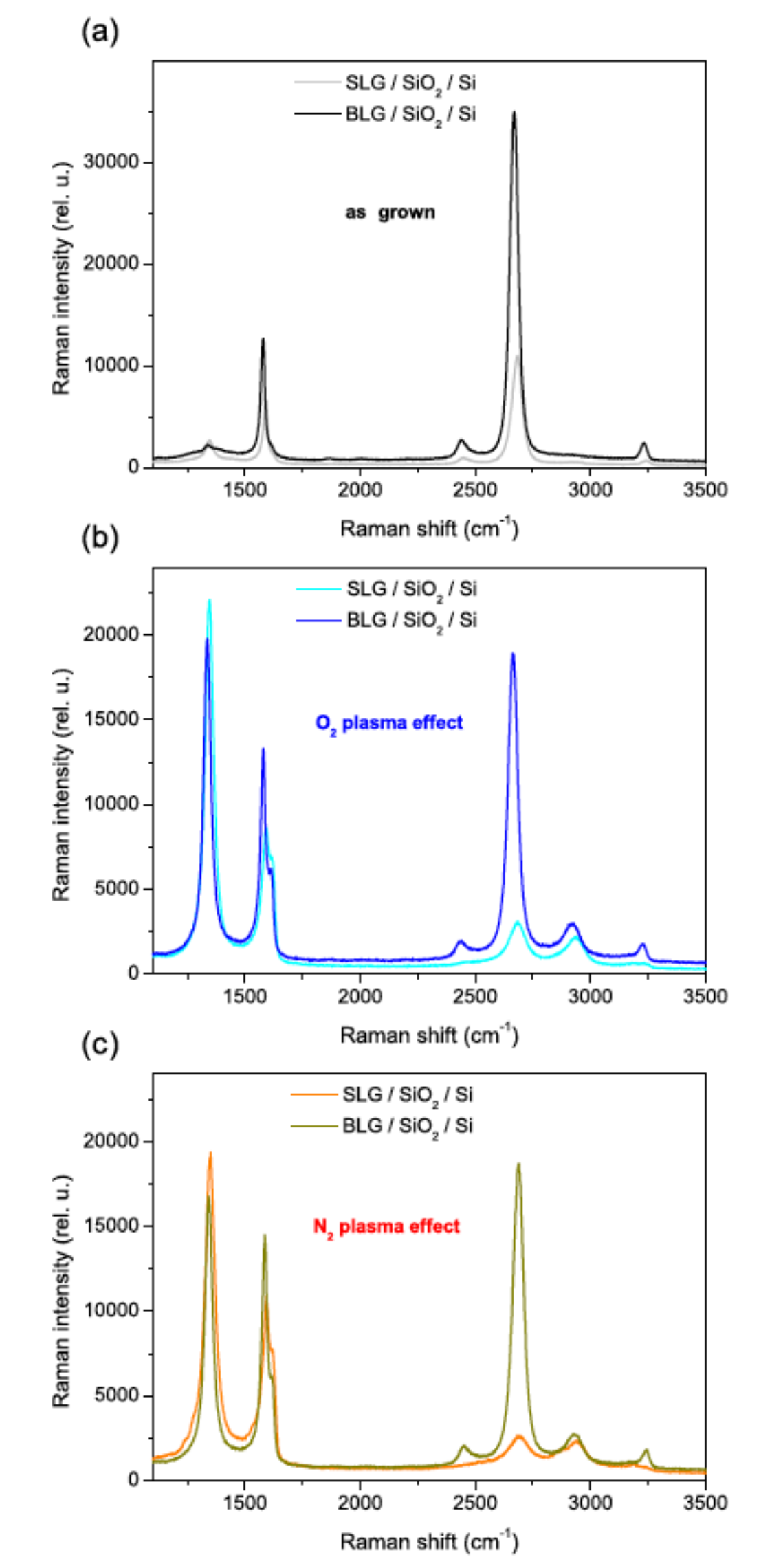}
\end{center}
\caption{Typical room-temperature Raman spectra of CVD grown SLG and BLG (a) as-transferred onto the SiO$_2$(300 nm)/Si(100) substrate, (b) exposed to O$_2$ plasma, and (c) exposed to N$_2$ plasma. The laser excitation wavelength is 514.5 nm.}
\label{Fig3}
\end{figure}

The as-transferred original and exposed to O$_2$ or N$_2$ plasma SLG and BLG samples were characterized by Raman spectroscopy. Their representative Raman spectra are shown in figures 3(a-c). The predominant D, G, and 2D features observed in the RS at about 1320-1350, 1570-1584, and 2640-2680 cm$^{-1}$, respectively, have been discussed widely in the literature [5,6,47-51]. According to the earlier studies, the G band originates from the first-order RS and is associated with the doubly degenerate zone-center $E_{\rm 2g}$ phonon mode. The D and 2D bands are due to the first- and second-order zone-boundary phonons, respectively. However, contrary to the 2D band, the D mode requires defects to be activated. In addition, the D' band may appear at 1602-1625 cm$^{-1}$ arising from the intravalley defect-induced double-resonance process. Therefore, an increase in the intensity of the D and D' peaks associated with phonon scattering at defect sites indicates the presence of defects. The  D peak is relatively weakly pronounced in the Raman spectra measured on the studied original SLG and BLG samples and the D' peak can hardly be distinguished there as well, which indicates the insignificant imperfectness of the CVD grown graphene (figure 3(a)). This is further supported by fairly narrow widths of the observed Raman bands. Note that the 2D peak in the original BLG sample is located at lower wavenumbers than in the original SLG sample (see figure 3(a)). This is in contrast to the situation expected for $n$-layer graphene, where the 2D peak shifts to higher wavenumbers with increasing the number of layers [3]. In our case, the observed downshift of the 2D peak from 2682 cm$^{-1}$ in SLG to 2669 cm$^{-1}$ in BLG (see figure 3(a), with more details presented in figures S3(a) and (d) of the Supplementary online information) indicates that bilayer CVD graphene has a quasi-free-suspended top layer. Indeed, it is reported that the two-phonon 2D mode of the free-standing graphene monolayers is downshifted in frequency compared to that of the supported regions [52]. The downshift can be also noticed for the G, D and D' peak positions in original BLG with respect to their positions in original SLG (see figures S3(a) and (d) of the Supplementary online information).

In addition, from the previous studies, it is established that the intensity ratio of the D and G bands ($I_{\rm D}$/$I_{\rm G}$) is inversely proportional to the in-plane cluster size $L_a$ in poly- and nanocrystalline graphites: $I_{\rm D}$/$I_{\rm G}$=$C(\lambda)/L_a$, where $C$(514.5 nm) $\approx$4.4 nm [50,51,53,54]. Evidently, the introduction of vacancies and O- or N-related defects into the graphene lattice will lead to the smaller crystallite size. From the Raman spectrum of as-transferred CVD SLG shown in figure 3(a), we have $I_{\rm D}$/$I_{\rm G}$ = 0.41 and the estimated in-plane cluster size $L_a$ $\approx$ 11 nm. In O$_2$ and N$_2$ plasma treated SLG (see figures 2(b) and (c)) we obtain  noticeably higher $I_{\rm D}$/$I_{\rm G}$ values of 2.4 and 1.7, and the estimated $L_a$ sizes are $\approx$ 1.8 and 2.6 nm, respectively. The $I_{\rm D}$/$I_{\rm G}$ intensity ratio for doped graphene increases, and the D' peak becomes clearly pronounced. This gives evidence for a sizable amount of defects introduced by O$_2$ and N$_2$ plasma treatment in SLG, which most probably can be associated with defective structures containing carbon vacancies and epoxides (for O$_2$ plasma treatment) or pyridinic (for N$_2$ plasma treatment) functional groups. In addition, the intensity ratio $I_{\rm 2D}$/$I_{\rm G}$ has been used to characterize the doping level. Indeed, it has been reported that $I_{\rm 2D}$/$I_{\rm G}$ depends on the electron or hole densities [55]. In SLG exposed to the O$_2$ and N$_2$ plasma treatment (see figures 3(b) and (c)), we have $I_{\rm 2D}$/$I_{\rm G}$ values of 0.36 and 0.24, respectively, indicating that the doping level is higher than 4 $\times$ 10$^{13}$ cm$^{-2}$ [55], that is of about 10\%. On the other hand, one can see that on the quasi-free-suspended upper BLG sheets (see figure 3(a-c)), the 2D-mode intensity is significantly enhanced, while the G mode is only slightly higher than that in SLG supported on an oxide-coated silicon substrate.
Recent theoretical studies suggest that an enhanced intensity of the 2D mode in doped graphene results from reduced electron-electron scattering [56]. One can notice that the D mode measured on the upper BLG sheet exposed to O$_2$ or N$_2$ plasma treatment has a lower intensity than that in O$_2$ or N$_2$ treated SGL. This can imply somewhat lower introduced defect density in the plasma treated BLG samples. Moreover, it should be noted that an additional disorder can come in BLG from ripples, edges of the perforated graphene structure, and other defects. 

\begin{figure}
\begin{center}
\hspace{2.50cm}
\includegraphics*[width=130mm]{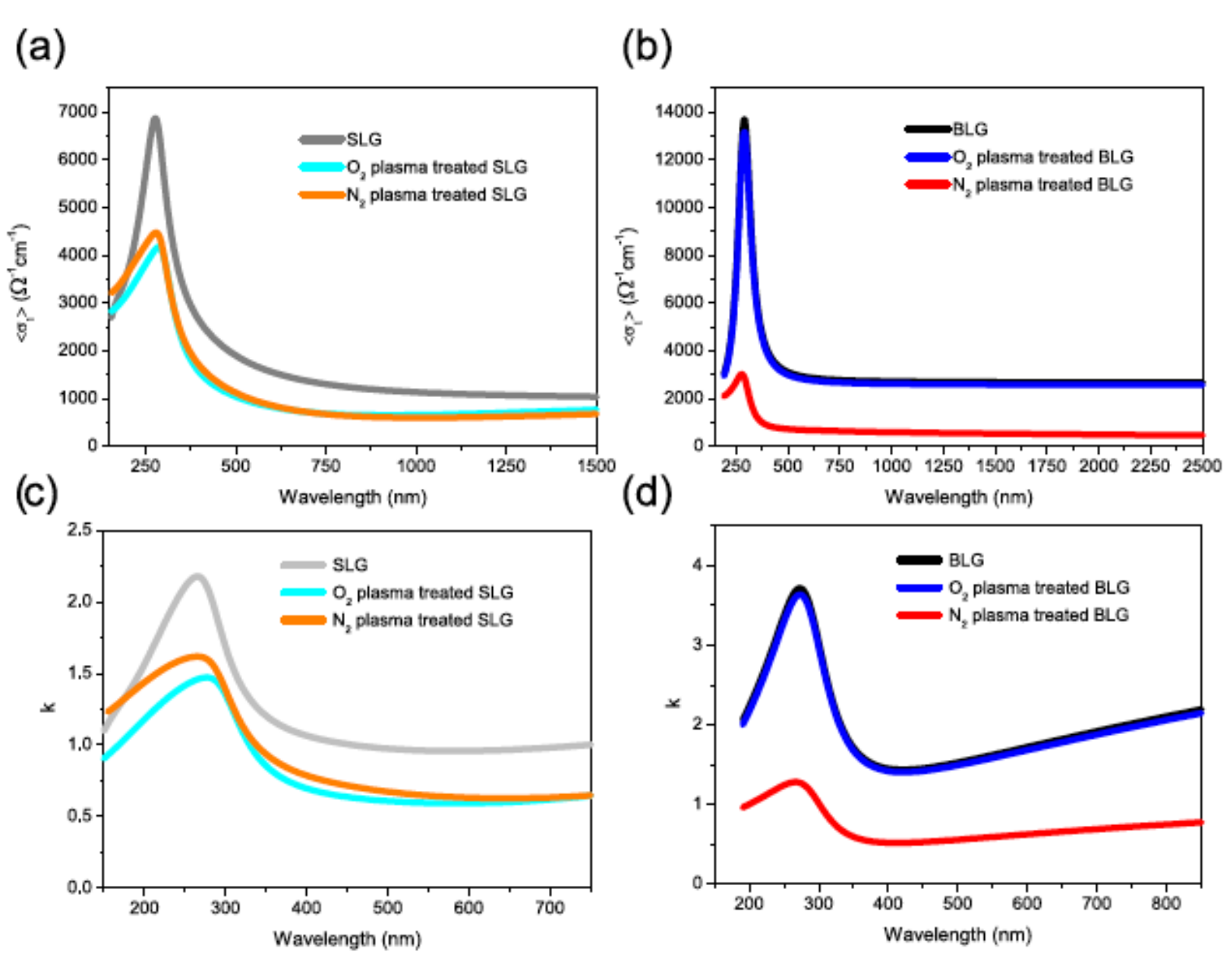}
\end{center}
\caption{(a) Simulated optical conductivity (in EMA) of as-transferred and exposed to O$_2$ or N$_2$ plasma SLG. (b) The simulated optical conductivity (in EMA) of the upper sheet of as-transferred and exposed to O$_2$ or N$_2$ plasma BLG. The extinction coefficient k for as-transferred and exposed to O$_2$ or N$_2$ plasma (c) SLG and (d) BLG.}
\label{Fig4}
\end{figure}

Figures 4(a) and (b) present the optical conductivity for as-transferred and exposed to O$_2$ or N$_2$ plasma SLG and BLG, respectively, simulated by multilayer models in EMA from the ellipsometric angles measured in the wide spectral range using the J.A. Woollam VASE software [57]. In accordance with the theoretical and experimental studies [58-60], the effective optical conductivity spectrum, $\left\langle \sigma_1(\lambda) \right\rangle$, obtained for as-transferred SLG (see figure 4(a)) exhibits several key features. In the near-IR spectral range of 830-2500 nm the optical conductivity associated with Dirac fermions is well described by the universal value $\pi e^2/2h$, where $e$ is the electron charge, $c$ is the speed of light, and $h$ is Planck's constant [59,60]. With decreasing wavelength $\left\langle \sigma_1 \right \rangle$ steadily grows in the visible spectral range, while in the UV range it exhibits the pronounced peak at $E_{ex}$ = 277 nm (4.5 eV). This feature arises from the interband transition in graphene from the bonding to the antibonding $\pi$ state near the saddle-point singularity at the M point of the Brillouin zone [58]. The observed excitonic band has an asymmetric line shape with higher absorption at the longer wavelengths, which is attributed to strong attractive electron-hole interactions of the quasiparticles near the 2D saddle point singularity. We found that the exciton peak noticeably decreases as a result of O$_2$ or N$_2$ plasma treatment, and the whole excitation band becomes strongly renormalized exhibiting pseudogap-like behavior near 800 nm. These effects may be associated with screening of additional charges introduced by plasma treatment. Figure 4(b) shows the effective optical conductivity $\left\langle \sigma_1 \right\rangle$ for the upper sheet of as-transferred and exposed to O$_2$ or N$_2$ plasma BLG. It follows from the figure that N$_2$ plasma has a pronounced effect on $\left\langle \sigma_1 \right\rangle$ of the upper BLG sheet and its effective optical conductivity is strongly suppressed in the whole investigated spectral range. In addition, the EMA simulation indicates higher porosity in the morphology of the upper BLG sheet treated with N$_2$ plasma in comparison to the as-transferred upper BLG sheet. According to the present AFM study, the applied oxygen plasma conditions leads to weaker etching of the upper BLG sheet than nitrogen plasma (see figure 2(d-f) and S2(a-c)). However, the effect of O$_2$ plasma on $\left\langle \sigma_1\right\rangle$ of the upper BLG sheet cannot be seen from figure 4(b). We suggest that this can be explained by somewhat different morphology of the original and O$_2$ plasma treated BLG samples. Indeed, the effective optical conductivity $\left\langle \sigma_1 \right\rangle$ of the as-transferred upper BLG sheet is approximately two times larger in the near-IR spectral range than that of the as-transferred SLG sheet, which indicates its complex morphology, as it was investigated by the present AFM study (see figures 2(a-c)). Moreover, the extinction coefficient ($k$) is usually used to characterize the plasma damage effect [61]. Figures 4(c) and (d) show the extinction coefficient for as-transferred and exposed to O$_2$ or N$_2$ plasma SLG and BLG, respectively, obtained from our spectroscopic ellipsometry analysis. In particular, our results for as-transferred SLG are in a fair agreement with those of the previous work where the complex refractive index of large-area polycrystalline CVD graphene was studied by spectroscopic ellipsometry [62]. The extinction coefficient obtained for the visible region  $k$ $\sim$ 1.3 is close to the value of bulk graphite [63]. Deviation of the $k$ values for the as-transferred upper BLG sheet signifies its complex morphology. Similarly, as for $\left\langle \sigma_1 \right\rangle$, N$_2$ plasma has a pronounced effect on $k$ of the upper BLG sheet, where it is strongly suppressed in the whole investigated spectral range.

Figure 5 shows room-temperature PL emission and PL excitation (PLE) spectra of SLG/SiO$_2$/Si and BLG/SiO$_2$/Si. In the PLE measurements, the spectra were monitored near to the emission peaks. The samples of as-transferred SLG and BLG reveal the similar strongly pronounced PL emission band peaking near 390 nm under the 250 nm excitation. The PL band around 390 nm was also observed from the SiO$_2$/Si substrate (for more details see supplementary online information). The present PL measurements indicate that there might be a 10--15\% increase in the 390 nm PL intensity for the SLG samples exposed to O$_2$ or N$_2$ plasma treatment (see figure 5(a)). In addition, we found that the SLG samples exposed to O$_2$ or N$_2$ plasma exhibits noticeably enhanced PL around 450 nm (see also figures S6(b) and (d) and insets). The PL emission around 450 nm was found to be more pronounced in the N$_2$ plasma treated SLG samples subjected to the stronger N$_2$ plasma conditions. The PL signal from the BLG sample subjected to O$_2$ plasma treatment is very weak (see also figure S6(f) and inset). In contrast, the observed PL caused by N$_2$ plasma treatment of the BLG sample is by far more pronounced and exhibits clearly different spectral properties. Under the 250 nm excitation, another strongly pronounced emission band peaking at 475 nm appears in addition to the intense PL at 390 nm. Moreover, the observed PL emission is extended far to longer wavelengths up to 800 nm (see figure 5(b)). 
\begin{figure}
\begin{center}
\hspace{2.50cm}
\includegraphics*[width=130mm]{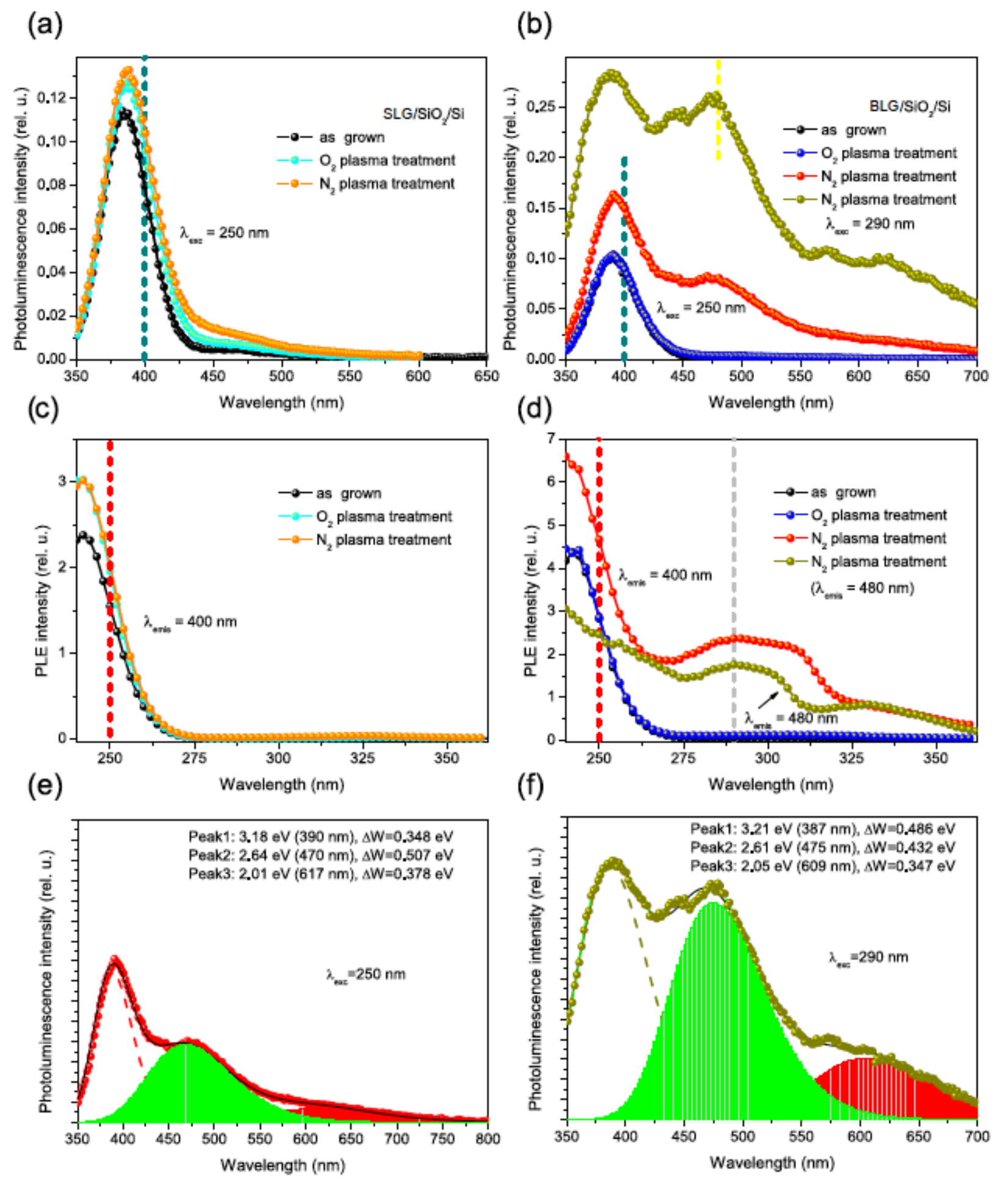}
\end{center}
\caption{The room-temperature (a,b) PL emission (the y-axes show the same rel.u.) and (c,d) PLE spectra of as-transferred and exposed to O$_2$ or N$_2$ plasma SLG/SiO$_2$/Si and BLG/SiO$_2$/Si (the y-axes show the same rel.u.). (e,f) The PL emission spectra (under 250 and 290 nm light excitation, respectively) of BLG/SiO$_2$/Si exposed to N$_2$ plasma (corresponding to the PL data shown in (b)), represented by a the sum of the contributing Gaussian bands. }
\label{Fig5}
\end{figure}

In the PLE spectrum registered for the UV emission at 400 nm, there is an excitation band covering the region from 270 nm to 320 nm (see figure 5(d)). It is not seen in the PLE spectrum of the SLG\ samples, being exclusive for N$_2$ plasma treated BLG, as follows from figures 5(c) and (d). The PLE band attains its maximum at the spectral position of the strong exciton peak $E_{ex}$ $\sim$288 nm (4.3 eV) in the optical conductivity spectrum (see figure 4(b)). The PLE band intensity monotonously changes with increasing wavelength showing flat featureless behavior in the spectral range from 288 to 310 nm followed by the sharp intensity decrease (figure 5(d)). Figures 5(e) and (f) show the PL spectra of N$_2$ plasma treated BLG obtained under 250 and 290 nm light excitation, respectively, which are represented by the sum of three main Gaussian bands at ~390, ~475, and ~610--620 nm (here the constant background is subtracted). We note that the PL signal presented in figures 5(b) and (e,f) contains contribution from the SiO$_2$/Si substrate. One can follow that an increase in the excitation wavelength up to 290 nm leads to the appreciable emission enhancement from N$_2$ plasma treated BLG, where the PL band peaking at ~475 nm becomes relatively more pronounced. As one can see from figure 5(d), the corresponding PLE spectrum registered for the green emission at 480 nm demonstrates the similar excitation band covering 275 nm to 310 nm peaking at the strong exciton peak position at $E_{ex}$ $\sim$288 nm (4.3 eV). These results support the evidence that the PLE band is directly related to the $\pi -\pi$* excitation in graphene arising from the interband transition near the saddle-point singularity at the M point of the Brillouin zone [58]. We note that the PLE band position, which is centered at $\sim$296 nm (4.2 eV), did not shift with increasing the PL emission wavelength from ~400 to ~480 nm, as it would be expected if the origin of the observed PL was related to GQDs [15]. We suggest that the discovered PLE band can be attributed to the  $\pi -\pi$* excitonic band tail states related to the  energy levels arising at numerous edges existing in perforated graphene, including the edges with carbon-nitrogen bonds. Then, the recombination of charge carriers involving these energy levels can lead to the  observed UV (~390 nm) and green (~475 nm) emission discovered in N$_2$ plasma treated BLG. Indeed, when graphene is placed into nitrogen plasma, carbon atoms can be partly replaced by nitrogen atoms at the edges of graphene, where pyridinic N, pyrrolic N, and quaternary N (or graphitic N) may exist inside the pristine graphene lattice [24].   The peculiar defect structure arising at the graphene edges will be determined by the applied N$_2$ plasma conditions. In particular, additional regularities in the structure of defects and/or intercalated nitrogen atoms can give rise to the zone folding and to the to the corresponding replica states in the electronic spectrum of graphene [64]. The associated effects can be relevant to a variety of systems, such as, for example, graphene nanosheets [65-68], and should be studied in more detail experimentally and theoretically. 

The PL emission found in the SLG samples exposed to O$_2$ or N$_2$ plasma, which is observed in the spectral range of about  450--460 nm, is comparatively weak. We suppose that the plasma treatment effect is different in SLG supported by substrate leading to doping in the graphene basal plane, which in turn may lead to changes of the defect structure and their PL properties in the near-interface region. The PL signal from BLG exposed to O$_2$ plasma treatment is also weak and significantly broadened. According to our present AFM study, the applied oxygen plasma conditions lead to weaker etching effect of the upper BLG sheet at room temperature. In is shown that efficient oxygen plasma etching occurs at elevated temperatures [46]. The defect edge structure will be different in O$_2$ treated graphene, where a very low amount of carboxylic acids (HCO$_2$) can be created.

\section{Conclusions}  

In summary, we found that BLG exposed to N$_2$ plasma exhibits efficient UV-green emission in contrast to a relatively weak UV PL signal registered in SLG exposed to the same N$_2$ plasma conditions. Our AFM study demonstrated that the upper sheet of BLG has a complex morphology structure peculiar for bilayer CVD graphene having a quasi-free-suspended top layer due to intercalated water or wet chemicals trapped between the graphene layers. In addition, the observed downshifts of the main Raman modes indicate that the CVD-grown BLG sample has a quasi-free-suspended top layer. The top layer becomes partially etched during exposure to N$_2$ plasma, and the perforated graphene is produced. In this case, nitrogen plasma processing at the edges of the perforated graphene is more efficient than doping in the graphene basal plane, the latter being, seemingly, more pertinent to SLG/SiO$_2$/Si. The PL spectra of BLG exposed to N$_2$ plasma show three main emission bands with the peaks at 390, 475, and 610--620 nm. In the PLE spectrum, there is a pronounced excitation band, which turns out to be in the range of the strong exciton peak $E_{ex}$ $\sim$288 nm (4.3 eV) in the optical conductivity. We suggest that the discovered PLE band can be attributed to the  $\pi -\pi$* excitonic band tail states arising due to the existence of numerous edges in perforated graphene. Then, the observed green PL emission can be associated with the edge states in perforated graphene filled with nitrogen [66]. The nitrogen plasma processing of bilayer CVD graphene with the formation of the peculiar defect edge structure may have a great potential for the development of a new class of advanced optoelectronic materials.

\section*{Acknowledgments}
This work was supported by the MEYS LO1409 project funded by the Infrastructure SAFMAT LM2015088, operational Program Research, Development and Education, co-financed by the European Structural and Investment Funds and the state budget of the Czech Republic within Center of Advanced Applied Sciences, Reg. No. 
CZ.02.1.01/0.0/0.0/16$_-$019/0000778, project SOLID21 
Reg. No.  CZ.02.1.01/0.0/0.0/16$_-$019/0000760), 
and partially supported by the Basic Science Research Program (2017R1D1A1B03035102 and 2017R1D1A1B03032759), the International Research and Development program (2016R1A6A1A03012877) through the NRF and MEST of Korea, by Russian Foundation for Basic Research according to the research project N19-29-03050. NNK is grateful to the Royal Society grant for the support and Loughborough University for the hospitality.. The authors are grateful to Alexandr Gorbatsevich and Mikhail Skorikov for helpful discussions.  The work of F.V.K. was supported by the Government of the Russian Federation through the ITMO Professorship Program.

\newpage
 
\section*{References}

\hspace{0.8cm}[1] Geim A K and Novoselov K S 2007 {\it Nat. Mater.} {\bf 6} 183

[2] Geim A K 2009 {\it Science} {\bf 324} 1530

[3] Zhu Y W, Murali S, Cai W W, Li X S, Suk J W, Potts J R and Ruoff R S, 2010 {\it Adv. Mater.} {\bf 22} 3906

[4] Nair R R {\it et al.} 2010 {\it Small} {\bf 6} 2877

[5] Gokus T, Nair R R, Bonetti A, Bohmler M, Lombardo A, Novoselov K S, Geim A K, Ferrari A C and Hartschuh A 2009 {\it ACS Nano} {\bf3} 3963

[6] Nourbakhsh A, Cantoro M, Vosch T, Pourtois G, Clemente F, Veen M H V D, Hofkens J, Heyns M M, Gendt S D and Sels B F 2010 {\it Nanotechnology} {\bf 21} 435203

[7] Wang H, Maiyalagan T and Wang X 2012 {\it ACS Catalysis} {\bf 2} 781

[8] Smith D {\it et al.} 2015 {\it ACS Nano} {\bf 9} 8279

[9] Liu F, Choi J Y and Seo T S 2010 {\it Biosen. Bioelectron.} {\bf 25} 2361

[10] Loh K P, Bao Q, Eda G and Chhowalla M 2010 {\it Nat. Chem.} {\bf 2} 1015

[11] Ponomarenko L A, Schedin F, Katsnelson M I, Yang R, Hill E W, Novoselov K S and Geim A K 2008 {\it Science} {\bf 320} 356

[12] Li X, Wang X, Zhang L, Lee S and Dai H 2008 {\it Science} {\bf 319} 1229

[13] Li Y, Hu Y, Zhao Y, Shi G, Deng L, Hou Y and Qu L 2011 {\it Adv. Mater.}  {\bf 23} 776

[14] Peng J {\it et al.} 2012 {\it Nano Lett.} {\bf 12} 844

[15] Zhu S {et al.} 2012 {\it Adv. Funct. Mater.} {\bf 22} 4732

[16] Geng X {\it et al.} 2010 {\it Adv. Mater.} {\bf 22} 638
 
[17] Son D I, Kwon B W, Park D H, Seo W, Yi Y, Adgadi B, Lee C and Choi W K 2012 {\it Nature Nanotech.} {\bf 7} 465

[18] Kim J K, Park M J, Kim S J, Wang D H, Cho S -P, Bae S, Park J H and Hong B H 2013 {\it Acs Nano} {\bf7} 7208

[19] Bkakri R, Kusmartseva O E, Kusmartsev F V, Song M, Sfaxi L and Bouazizi A 2015 {\it Synth. Met.} {\bf 203} 74

[20] Bkakri R, Kusmartseva O E, Kusmartsev F V, Song M and Bouazizi A 2015 {\it J. Lumin.} {\bf 161} 264

[21] Hummer W and Offeman R 1958 {\it J. Am. Chem. Soc.} 1339 

[22] Li X, Zhang G, Bai X, Sun X, Wang X, Wang E, Dai H 2008 {\it Nat. Nanotechnol.}  {\bf 3} 538

[23] Hirata M, Gotou T, Horiuchi S, Fujiwara M and Ohba M 2004 {\it Carbon}  {\bf 42} 2929

[24] Geng D, Yang S, Zhang Y, Yang J, Liu J, Li R, Sham T -K, Sun X, Ye S and Knights S 2011 {\it Appl. Surf. Sci.} {\bf 257} 9193

[25] Wei D, Liu Y, Wang Y, Zhang H, Huang L and Yu G 2009  {\it Nano Lett.} {\bf 9} 1752 

[26] Luo Z, Lim S, Tian Z, Shang J, Lai L, MacDonald B, Fu C, Shen Z, Yu T and Lin J 2011 {\it J. Mater. Chem.} {\bf 21} 8038 

[27] Jafri R I, Rajalakshmi N and Ramaprabhu S 2010 {\it J. Mater. Chem.}  {\bf 20} 7114 

[28] Shao Y, Zhang S, Engelhard M H, Li G, Shao G, Wang Y, Liu J, Aksay I A and Lin Y 2010 {\it J. Mater. Chem.} {\bf 20} 7491

[29] Wang Y, Shao Y, Matson D W, Li J and Lin Y 2010 {\it ACS Nano} {\bf 4} 1790

[30] Wang X, Li X, Zhang L, Yoon Y, Weber P K, Wang H, Guo J, Dai H 2019 {\it Science} 324 768

[31] Panchakarla L S, Subrahmanyam K S, Saha S K, Govindaraj A, Krishnamurthy H R, Waghmare U V, and Rao C N R 2009 {Adv. Mater.} 21 4726

[32] Abdelkader A M and Fray D J 2017 {\it Nanoscale} 9 14548

[33] Fasolino A, Los J H and Katsnelson M I 2007 {\it Nat. Mater.} {\bf 6} 858

[34] Meyer J C, Geim A K, Katsnelson M I, Novoselov K S, Booth T J and Roth R 2007 {\it Nature} {\bf 446} 60

[35] O'Hare A, Kusmartsev F V and Kugel K I 2012 {\it Nano Lett.} {\bf 12} 1045

[36] Ben Gouider Trabelsi A, Kusmartsev F V, Kusmartseva O E, Robonson B and Kolosov O 2014 {\it Nanotechnology} {\bf 25} 165704

[37] Ben Gouider Trabelsi A, Kusmartsev F V, Forrester D M, Kusmartseva O E, Gaifullin M, Cropper P and Oueslati M 2016 {\it J. Mater. Chem. C} {\bf 4} 5829

[38] Ben Gouider Trabelsi A, Ouerghi A, Kusmartseva O E, Kusmartsev F V and Oueslati M 2013 {\it Thin Solid films} {\bf 539} 377

[39] Ben Gouider Trabelsi A, Kusmartsev F V, Gaifullin M B, Forrester D M, Kusmartseva A and Oueslati M 2017 {\it Nanoscale}  {\bf 9} 11463

[40] Teague M L, Lai A P, Velasco J, Hughes C R, Beyer A D, Bockrath M W, Lau C N and Yeh N C 2009 {\it Nano Lett.} {\bf 9} 2542

[41] Xu K, Cao P and Heath J R 2009 {\it Nano Lett.} {\bf 9} 4446

[42] Haddon R C 1993 {\it Science} {\bf 261} 1545

[43] Gulseren O, Yildirim T and Ciraci S 2001 {\it Phys. Rev. Lett.}  {\it 87} 116802

[44] Niyogi S, Hamon M A, Hu H, Zhao B, Bhowmik P, Sen R, Itkis M E and Haddon R C 2002 {\it Acc. Chem. Res.} {\bf 35} 1105

[45] Magnozzi M 2017 {\it Carbon} {\bf 118} 208

[46] Liu L, Sunmin R, Tomasik M R, Stolyarova E, Jung N, Hybertsen M S, Steigerwald M L, Brus L E and Flynn G W 2008 {\it Nano Lett.}  {\bf 8} 1965

[47] Ferrari A C {\it et al.} 2006 {\it Phys. Rev. Lett.} {\bf 97} 187401

[48] Cancado L G, Pimenta M A, Neces B R A, Dantas M S S and Jorio A 2004
{\it Phys. Rev. Lett.}  {\bf 93} 247401

[49] Dresselhaus M S, Dresselhaus G, Saito R and Jorio A 2005 {\it Phys. Rep.} {\bf 409} 47

[50] Tuinstra F and Koenig J L 1970 {\it J. Chem. Phys.}  {\bf 53} 1126

[51] Nemanich R J and Solin S A 1979 {\it Phys. Rev. B} {\bf 20} 392

[52] Berciaud S, Ryu S, Brus L E, and Heinz T F 2009 Nano Lett. 9 346

[53] Knight D S and White W B 1989  {\it J. Mater. Res.}  {\bf 4} 385

[54] Zhang C, Fu L, Liu N, Liu M, Wang Y and Liu Z 2011 {\it Adv. Mater.} {\bf 23} 1020

[55] Das A {\it et al.} 2008 {\it Nat. Nanotechnol.} {\bf 3} 210

[56] Basko D M 2008 Phys. Rev. B 78 125418

[57] Woollam J A, {\it VASE Spectroscopic Ellipsometry Data Acquisition and Analysis Software} 2010 J. A. Woollam Co.: Lincoln, NE, USA

[58] Mak K F, Shan J, Heinz T F 2011 {\it Phys. Rev. Lett.} {\bf 106} 046401

[59] Peres N M R, Guinea F and Castro Neto A H 2006 {\it Phys. Rev. B} {\bf 73} 125411

[60] Gusynin V P, Sharapov S G and Carbotte J P 2006 {\it Phys. Rev. Lett.}  {\bf 96} 256802

[61] Baklanov M R, de Marneffe J -F, Shamiryan D, Urbanowicz A M, Shi H, Rakhimova T V, Huang H and Ho P S 2013 {\it J. Appl. Phys.} {\bf 113} 041101

[62] Nelson F J, Kamineni V K, Zhang T, Comfort E S, Lee J U and Diebold A C 2010 {\it Appl. Phys. Lett.} {\bf 97} 253110

[63] Ni Z H, Wang H M, Kasim J, Fan H M, Yu T, Wu Y H, Feng Y P and Shen Z X 2007 {\it Nano Lett.} {\bf 7} 2758

[64] Lin Y, Chen G, Sadowski J T, Li Y, Tenney S A, Dadap J I, Hybertsen M S and Osgood R M 2019 {\it Phys. Rev. B} 2019, 99, 035428

[65] Liu F, Jang M -H, Ha H D, Kim J -H, Cho Y -H and Seo T S 2013 {\it Adv. Mater.} {\bf 25} 3657

[66] Song X H, Feng L, Deng S L, Xie S Y and Zheng L S 2017 {\it Adv. Materials Interfaces}  {\bf 15} 1700339

[67] Liu Y, Tiwari R P, Brada M, Bruder C, Kusmartsev F V, Mele E J 2015 {\it Phys. Rev. B} {\bf 92} 235438

[68] Liu Y, Brada M, Mele E J, Kusmartsev F V 2014 {\it Ann. Phys.} {\bf 526} 449

\end{document}